%% file: acl.tex
\pdfoutput=1

\documentclass[11pt]{article}

\usepackage[]{acl}

\usepackage{times}
\usepackage{latexsym}

\usepackage[T1]{fontenc}

\usepackage[utf8]{inputenc}

\usepackage{microtype}

\usepackage{multicol}
\usepackage{booktabs}
\usepackage{makecell}
\usepackage{amsmath}
\usepackage{multirow}
\usepackage{mathtools}
\usepackage{xcolor}
\usepackage{enumitem}
\usepackage{subcaption}
\usepackage{float}
\usepackage{graphicx}
\usepackage{caption} 

\title{Augmenting Document Representations for Dense Retrieval \\with Interpolation and Perturbation}

\author{Soyeong Jeong$^1$
        \quad Jinheon Baek$^2$
        \quad Sukmin Cho$^1$ 
        \quad Sung Ju Hwang$^2$
        \quad Jong C. Park$^1$\thanks{\hspace{0.2cm}Corresponding author} \\
        School of Computing$^1$ \quad Graduate School of AI$^2$ \\
        Korea Advanced Institute of Science and Technology$^1$$^,$$^2$\\ 
       \texttt{\{syjeong,nelllpic,park\}@nlp.kaist.ac.kr}$^1$ \\
       \texttt{\{jinheon.baek,sjhwang82\}@kaist.ac.kr}$^2$}

\begin{document}
\maketitle
\begin{abstract}
Dense retrieval models, which aim at retrieving the most relevant document for an input query on a dense representation space, have gained considerable attention for their remarkable success. Yet, dense models require a vast amount of labeled training data for notable performance, whereas it is often challenging to acquire query-document pairs annotated by humans. To tackle this problem, we propose a simple but effective \textbf{D}ocument \textbf{A}ugmentation for dense \textbf{R}etrieval (DAR) framework, which augments the representations of documents with their interpolation and perturbation. We validate the performance of DAR on retrieval tasks with two benchmark datasets, showing that the proposed DAR significantly outperforms relevant baselines on the dense retrieval of both the labeled and unlabeled documents.

\end{abstract}

\input{1introduction}
\input{2relatedwork}
\input{3method}
\input{4experiments}
\input{5conclusion}

\vspace{-0.05in}
\section*{Acknowledgements}
\vspace{-0.1in}
This work was supported by Institute for Information and communications Technology Promotion (IITP) grant funded by the Korea government (MSIT) (No. 2018-0-00582, Prediction and augmentation of the credibility distribution via linguistic analysis and automated evidence document collection).

\section*{Ethical Statements}
\vspace{-0.05in}
Retrieving the most relevant documents from the user's query is increasingly important in a real-world setting, as it is widely used from web search, to question answering, to dialogue generation systems. Notably, our work contributes to the accurate retrieval of documents with the proposed data augmentation strategies, thus improving the document retrieval performances on real-world applications. However, we have to still consider the failure of retrieval systems on low-resource but high-risk domains (e.g., biomedicine), where the labeled data for training retrieval models is limited yet one failure can yield a huge negative impact. While we strongly believe that our data augmentation strategies -- interpolation and perturbation of document representations -- are also helpful to improve the retrieval performances on such low-resource domains, the model's prediction performance is still far from perfect, and more efforts should be made to develop a reliable system.

\bibliography{anthology,custom}
\bibliographystyle{acl_natbib}

\input{6appendix}

\end{document}

%% file: 1introduction.tex
\section{Introduction}
Retrieval systems aim at retrieving the documents most relevant to the input queries, and have received substantial spotlight since they work as core elements in diverse applications, especially for open-domain question answering (QA)~\cite{Voorhees99odqa}. Open-domain QA is a task of answering the question from a massive amount of documents, often requiring two components, a retriever and a reader~\cite{ChenFWB17readerretriever, karpukhin2020dpr}. Specifically, a retriever ranks the most question-related documents, and a reader answers the question using the retrieved documents.

Traditional sparse retrieval approaches such as BM25~\cite{Robertson1994Okapi} and TF-IDF rely on term-based matching, hence suffering from the vocabulary mismatch problem: the failure of retrieving relevant documents due to the lexical difference from queries. To tackle such a problem, recent research focuses on dense retrieval models to generate learnable dense representations for queries and documents~\cite{karpukhin2020dpr}.


Despite their recent successes, some challenges still remain in the dense retrieval scheme for a couple of reasons. First, dense retrieval models need a large amount of labeled training data for a decent performance. However, as Figure~\ref{fig:motivation} shows, the proportion of labeled query-document pairs is extremely small since it is almost impossible to rely on humans for the annotations of a large document corpus. Second, in order to adapt a retrieval model to the real world, where new documents constantly emerge, handling unlabeled documents that are not seen during training should obviously be considered, but remains challenging.

\input{0motivation_fig}

To automatically expand the query-document pairs, recent work generates queries from generative models~\cite{liang2020embedding,ma2021zero} or incorporates queries from other datasets~\cite{qu2021rocketqa}, and then generates extra pairs of augmented queries and documents. However, these query augmentation schemes have serious and obvious drawbacks. First, it is infeasible to augment queries for every document in the dataset (see the number of unlabeled documents in Figure~\ref{fig:motivation}), since generating and pairing queries are quite costly. Second, even after obtaining new pairs, we need extra training steps to reflect the generated pairs on the retrieval model. Third, this query augmentation method does not add variations to the documents but only to the queries, thus it may be suboptimal to handle enormous unlabeled documents.

Since augmenting additional queries is costly, the question is then if it is feasible to only manipulate the given query-document pairing to handle numerous unlabeled documents. To answer this, we first visualize the embeddings of labeled and unlabeled documents. Figure~\ref{fig:motivation} shows that there is no distinct distributional shift between labeled and unlabeled documents. Thus it could be effective to manipulate only the labeled documents to handle the nearby unlabeled documents as well as the labeled documents. Using this observation, we propose a novel document augmentation method for a dense retriever, which not only interpolates two different document representations associated with the labeled query (Figure~\ref{fig:concept} (b)), but also stochastically perturbs the representations of labeled documents with a dropout mask (Figure~\ref{fig:concept} (c)). One notable advantage of our scheme is that, since it manipulates only the representations of documents, our model does not require explicit annotation steps of query-document pairs, which makes it highly efficient. We refer to our overall method as Document Augmentation for dense Retrieval (DAR).

\input{0concept_fig}

We experimentally validate our method on standard open-domain QA datasets, namely Natural Question (NQ)~\cite{Kwiatkowski2019NQ} and TriviaQA~\cite{Joshi2017TriviaQAAL} (TQA), against various evaluation metrics for retrieval models. The experimental results show that our method significantly improves the retrieval performances on both the unlabeled and labeled documents. Furthermore, a detailed analysis of the proposed model shows that interpolation and stochastic perturbation positively contribute to the overall performance. 

Our contributions in this work are threefold:
\vspace{-0.1in}
\begin{itemize}[itemsep=0.05mm, parsep=1pt, leftmargin=*]
  \item We propose to augment documents for dense retrieval models to tackle the problem of insufficient labels of query-document pairs.
  \item We present two novel document augmentation schemes for dense retrievers: interpolation and perturbation of document representations. 
  \item We show that our method achieves outstanding retrieval performances on both labeled and unlabeled documents on open-domain QA tasks.
\end{itemize}


%% file: 0motivation_fig.tex
\begin{figure}

\centering

\begin{subfigure}[b]{0.22\textwidth}
    \centering
    \includegraphics[width=\textwidth]{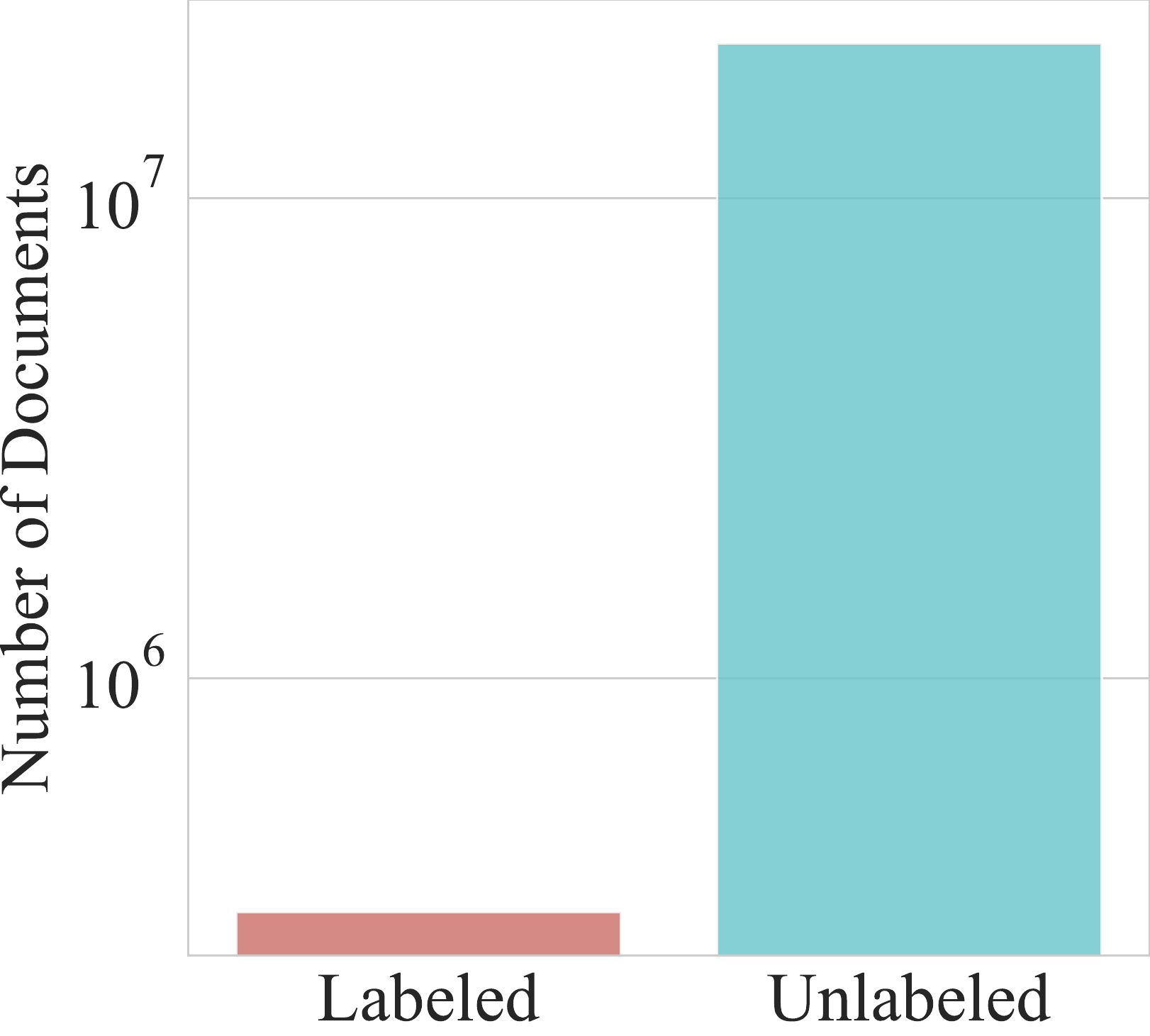}
    \label{fig:moti1}
\end{subfigure}
\hfill
\begin{subfigure}[b]{0.22\textwidth}
    \centering
    \includegraphics[width=\textwidth]{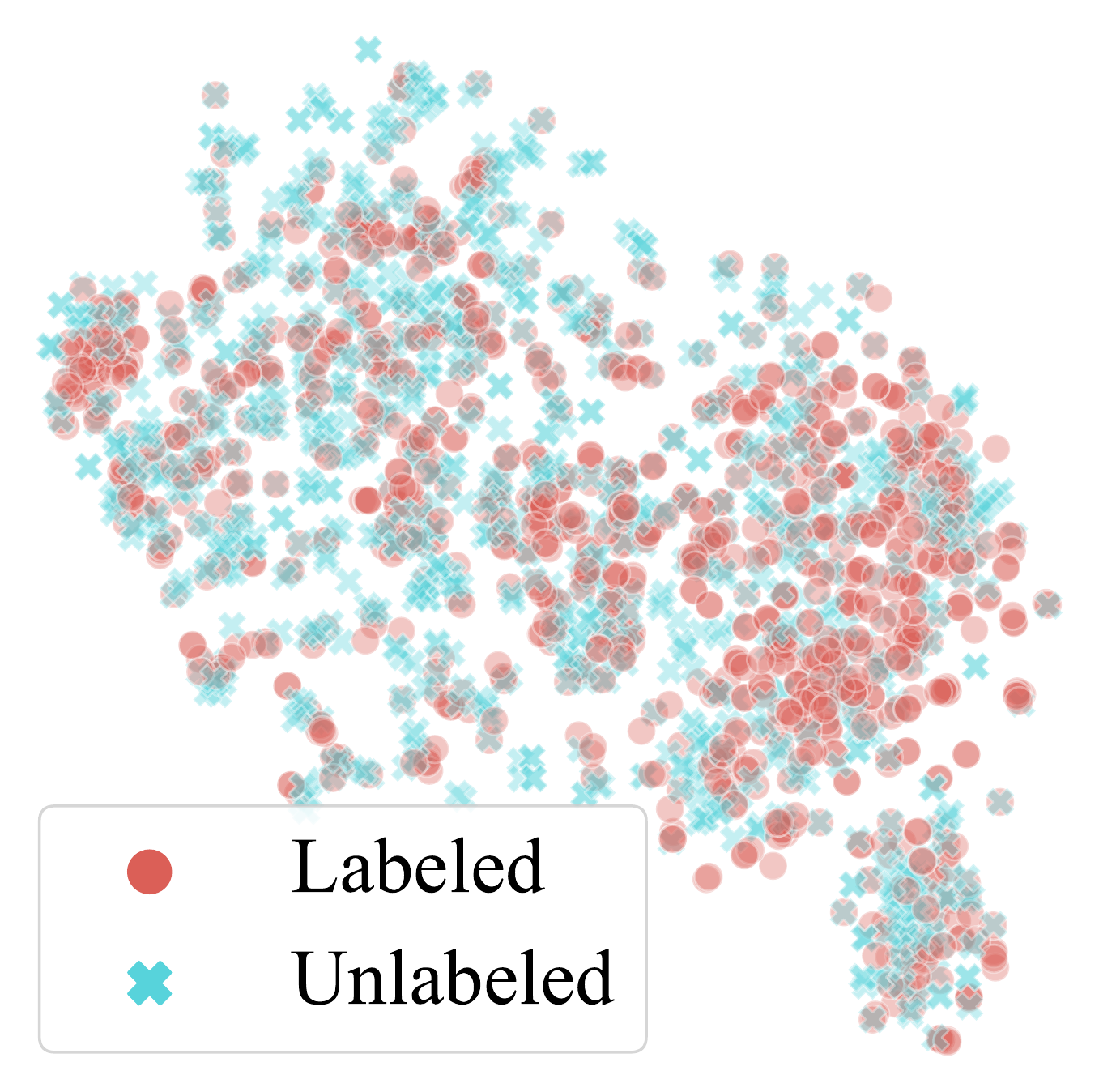}
    \label{fig:moti2}
\end{subfigure}
\vspace{-0.3in}
\caption{\small (Left) The number of labeled and unlabeled documents for the Natural Question dataset. (Right) T-SNE~\cite{tsne} visualization of randomly sampled document representations from the DPR model.}
\vspace{-0.2in}
\label{fig:motivation}
\end{figure}

%% file: 0concept_fig.tex
\begin{figure}[t!]
\begin{center}
\includegraphics[width=0.48\textwidth]{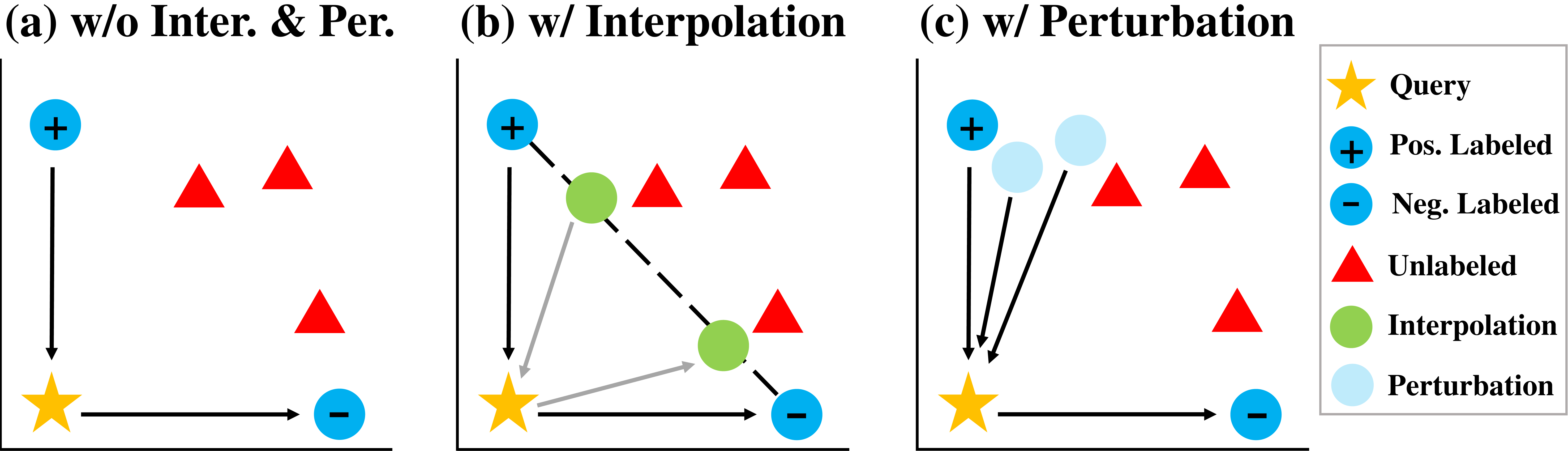}
\end{center}
\vspace{-0.15in}
\caption{
\small Our document augmenting schemes of interpolation and perturbation on a dense representation space. Pos. and Neg. denote positive and negative documents to the query.
}
\vspace{-0.2in}
\label{fig:concept}

\end{figure}

%% file: 2relatedwork.tex
\section{Related Work}

\paragraph{Dense Retriever}
Dense retrieval models~\cite{lee2019latentretrieval, karpukhin2020dpr} have gained much attention, which generate dense representations for queries and documents. However, dense retrieval faces a critical challenge from limited training data. Recent work has addressed such a problem by generating extra query-document pairs to augment those pairs to the original dense retrieval model~\cite{liang2020embedding,ma2021zero,qu2021rocketqa}, or by regularizing the model~\cite{axiomatic}. However, unlike ours that automatically augments data during a training phase, these methods require extensive computational resources for an additional generation step of explicitly query-document pairing before training the retriever.


\paragraph{Data Augmentation}
Since data augmentation is crucial to the performance of deep neural networks, it is widely applied to diverse domains~\cite{ShortenK19DAsurveyCV, hedderich2021DAsurveyNLP}, where interpolation and perturbation are dominant methods. Mixup interpolates two items, such as pixels of images, to augment the training data~\cite{zhang2018mixup, VermaLBNMLB19ManifoldMixup}, which is also adopted for NLP~\cite{chen2020mixtext, yin2021batchmixup}. However, none of the previous work has shown the effectiveness of mixup when applied to retrieval tasks. Besides interpolation,~\citet{wei2019eda} and~\citet{ma2019nlpaug} proposed perturbation over words, and \citet{lee2021learning} proposed perturbation over word embeddings. \citet{jeong2021unsupervised} and \citet{gao2021simcse} perturbed text embeddings to generate diverse sentences and to augment positive sentence pairs in unsupervised learning. In contrast, we address dense retrieval, perturbing document representations with dropout~\cite{srivastava14dropout} in a supervised setting with labeled documents. 





%% file: 3method.tex
\input{0main_table}
\section{Method}
\label{sec:method}

We begin with the definition of dense retrieval.

\paragraph{Dense Retrieval} 
Given a pair of query $q$ and document $d$, the goal of dense retrieval is to correctly calculate a similarity score between them from the dense representations $\boldsymbol{q}$ and $\boldsymbol{d}$, as follows:
\begin{equation}
\begin{aligned}
    &f(q, d) = \text{sim}(\boldsymbol{q}, \boldsymbol{d}), \\
    &\boldsymbol{q} = E_Q(q; \theta_q) \quad \text{and} \quad \boldsymbol{d} = E_D(d; \theta_d),
\end{aligned}
\label{eq:ir}
\end{equation}
where $f$ is a scoring function that measures the similarity between a query-document pair, $\text{sim}$ is a similarity metric such as cosine similarity, and $E_Q$ and $E_D$ are dense encoders for a query and document, respectively, with parameters $\theta = \left( \theta_q, \theta_d \right)$.

A dense retrieval scheme generally uses the negative sampling strategy to distinguish the relevant query-document pairs from irrelevant pairs, which generates an effective representation space for queries and documents. We specify a relevant query-document pair as $(q, d^+) \in \tau^+$, and an irrelevant pair as $(q, d^-) \in \tau^-$, where $\tau^+ \cap \tau^- = \emptyset$. The objective function is as follows:
\begin{equation}
    \min_{\theta} \!\! \sum_{(q, d^+) \in \tau^+} \! \sum_{(q, d^-) \in \tau^-} \!\! \mathcal{L}(f(q, d^+), f(q, d^-)), \!
\label{eq:negative}
\end{equation}
where a loss function $\mathcal{L}$ is a negative log-likelihood of the positive document. Our goal is to augment a set of query-document pairs, by manipulating documents with their interpolation or perturbation, which we explain in the next paragraphs.

\paragraph{Interpolation with Mixup}
As shown in interpolation of Figure~\ref{fig:concept}, we aim at augmenting the document representation located between two labeled documents to obtain more query-document pairs, which could be useful to handle unlabeled documents in the middle of two labeled documents. To achieve this goal, we propose to interpolate the positive and negative documents $(d^+, d^-)$ for the given query $q$, adopting mixup~\cite{zhang2018mixup}. Note that, since the input documents to the encoder $E_D$ are discrete, we use the output embeddings of documents to interpolate them, as follows:
\begin{equation}
    \tilde{\boldsymbol{d}} = \lambda \boldsymbol{d^+}  + (1-\lambda) \boldsymbol{d^-},
\label{eq:interpolation}
\end{equation}
where $\tilde{\boldsymbol{d}}$ is the mixed representation of positive and negative documents for the given query $q$, and $\lambda \in [0, 1]$. We then optimize the model to estimate the similarity $\text{sim}(\boldsymbol{q}, \tilde{\boldsymbol{d}})$ between the interpolated document and the query as the soft label $\lambda$ with a binary cross-entropy loss. The output of the cross-entropy loss is added to the original loss in equation~\ref{eq:negative}. One notable advantage of our scheme is that the negative log-likelihood loss in equation~\ref{eq:negative} maximizes the similarity score of the positive pair, while minimizing the score of the negative pair; thus there are no intermediate similarities between arbitrary query-document pairs. However, ours can obtain query-document pairs having soft labels, rather than strict positive or negative classes, by interpolating the positive and negative documents.


\input{0efficiency}

\paragraph{Stochastic Perturbation with Dropout}
In addition to our interpolation scheme to handle unlabeled documents in the space of interpolation of two labeled documents, we further aim at perturbing the labeled document to handle its nearby unlabeled documents as shown in Figure~\ref{fig:concept} (c). In order to do so, we randomly mask the representation of the labeled document, obtained by the document encoder $E_D$, with dropout, where we sample masks from a Bernoulli distribution. In other words, if we sample $n$ different masks from the distribution, we obtain $n$ different query-document pairs $\left\{ (\boldsymbol{q}, \boldsymbol{d}^+_i) \right\}_{i=1}^{i=n}$ from one positive pair $(\boldsymbol{q}, \boldsymbol{d}^+)$. By doing so, we augment $n$ times more positive query-document pairs by replacing a single positive pair $(q, d^+)$ in equation~\ref{eq:negative}. Moreover, since the document perturbation is orthogonal to the interpolation, we further interpolate between the perturbed positive document $\boldsymbol{d}^+_i$ and the negative document $\boldsymbol{d}^-$ for the given query in equation~\ref{eq:interpolation}, to augment a soft query-document pair from perturbation.


\input{0ablation_all}

\paragraph{Efficiency}
Data augmentation methods are generally vulnerable to inefficiency, since they need a vast amount of resources to generate data and to forward the generated data into the large language model. However, since our interpolation and perturbation methods only manipulate the already obtained representations of the documents from the encoder $E_D$, we don't have to newly generate document texts and also to forward generated documents into the model, which greatly saves time and memory (see Table~\ref{tab:efficiency}). We provide a detailed analysis and discussion of efficiency in \textbf{Appendix}~\ref{appendix:results:efficiency}.

%% file: 0main_table.tex
\aboverulesep=0ex
\belowrulesep=0ex

\begin{figure*}[ht]
    \begin{minipage}{0.74\linewidth}
        \vspace{-0.175in}
        \centering
        \begin{center}
        \resizebox{1.0\textwidth}{!}{
        \renewcommand{\arraystretch}{0.9}
        \begin{tabular}[t]{l|ccccccccccccc}
        \toprule
        \multicolumn{1}{c|}{} & \multicolumn{6}{c}{\textbf{Natural Questions (NQ)}} & \multicolumn{6}{c}{\textbf{TriviaQA (TQA)}} \\
        \cmidrule(l{2pt}r{2pt}){2-7} \cmidrule(l{2pt}r{2pt}){8-13} 
         & \textbf{MRR} & \textbf{MAP} & \textbf{T-100} & \textbf{T-20} & \textbf{T-5} & \textbf{T-1} & \textbf{MRR} & \textbf{MAP} & \textbf{T-100} & \textbf{T-20} & \textbf{T-5} & \textbf{T-1} \\
        \midrule
        \multirow{1}{*}{BM25}       & 32.46 & 20.78 & 78.25 & 62.94 & 43.77 & 22.11 & 55.28 & 34.85 & 83.15 & 76.41 & 66.28 & 46.30 \\
        \midrule
        \multirow{1}{*}{DPR}        & 39.55 & 25.61 & 83.77 & 72.94 & 54.02 & 27.45 & 44.29 & 27.24 & 80.50 & 71.07 & 57.74 & 33.63\\
        \multirow{1}{*}{DPR w/ QA}  & 40.00 & 24.93 & 83.46 & 72.13 & 55.46 & 27.67 & 46.27 & 28.08 & 80.76 & 71.88 & 59.14 & 35.90 \\
        \multirow{1}{*}{DPR w/ DA}  & 41.28 & 26.60 & 83.68 & 72.83 & 55.51 & 29.31 & 46.08 & 27.82 & 80.42 & 71.55 & 58.64 & 35.85\\
        \multirow{1}{*}{DPR w/ AR}  & 41.18 & 26.04 & 83.60 & 73.41 & 55.51 & 29.11 & 45.13 & 27.57 & 80.65 & 71.68 & 58.09 & 34.52\\
        \midrule
        \multirow{1}{*}{DAR (Ours)} & \textbf{42.92} & \textbf{27.12} & \textbf{84.18} & \textbf{75.04} & \textbf{57.62} & \textbf{30.42} & \textbf{47.32} & \textbf{28.70} & \textbf{81.30} & \textbf{72.66} & \textbf{59.88} & \textbf{36.94}\\
        \multirow{1}{*}{QAR (Ours)} & \textbf{43.09} & \textbf{27.64} & \textbf{84.21} & \textbf{74.76} & \textbf{57.51} & \textbf{31.25} & \textbf{47.21} & \textbf{29.00} & \textbf{80.91} & \textbf{72.12} & \textbf{59.94} & \textbf{36.92}\\
        \bottomrule
        \end{tabular}
        }
        \end{center}
        \vspace{-0.18in}
        \captionof{table}{\small Retrieval results on NQ and TQA datasets, including the variant of our model -- QAR: applying data augmentation techniques to queries instead of documents. BM25 is the sparse retrieval model, whereas others are dense retrieval models. The best model and the second best model among dense retrievers are denoted in \textbf{bold}, which we aim to improve in this work.}
        \label{tab:main}
    \end{minipage}
    \hfill
    \begin{minipage}{0.24\linewidth}
        \centering
        \vspace{0.05in}
        \centerline{\includegraphics[width=1.0\columnwidth]{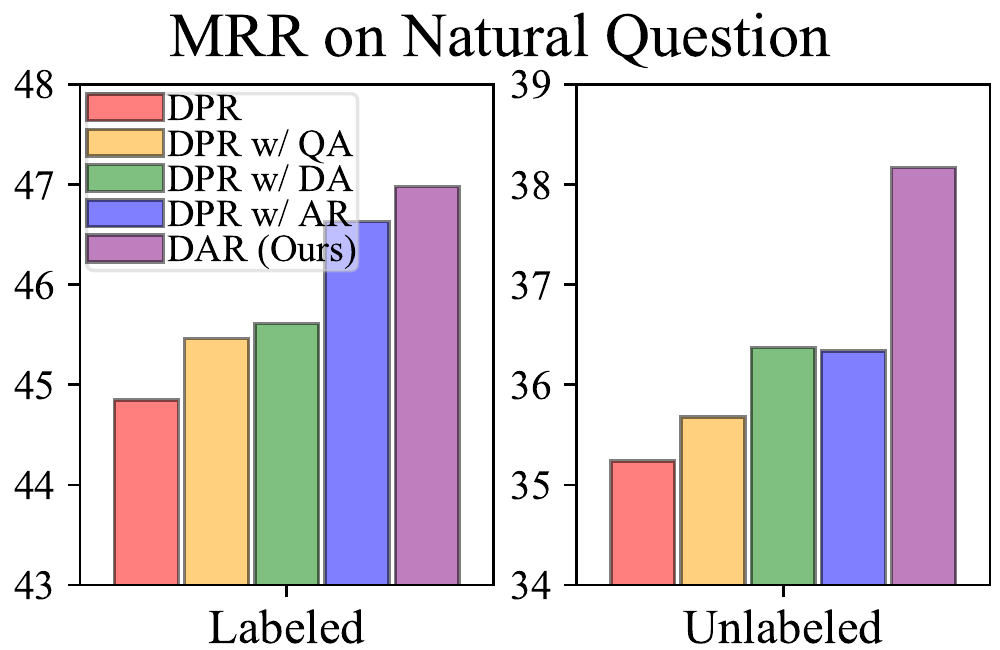}}
        \vspace{-0.1in}
        \caption{\small Retrieval results on the labeled and unlabeled documents in the NQ dataset with MRR as an evaluation metric.}
        \label{fig:seenunseen}
    \end{minipage}
    \vskip -0.175in
\end{figure*}

%% file: 0efficiency.tex
\aboverulesep=0ex
\belowrulesep=0ex

\begin{figure}[!t]
    \begin{minipage}{0.46\linewidth}
        \centering
        \begin{center}
        \resizebox{1.0\textwidth}{!}{
        \renewcommand{\arraystretch}{0.95}
        \begin{tabular}[h]{lc|cc}
        \toprule
        \multicolumn{2}{l|}{\textbf{\# Query}} & \textbf{MRR} & \textbf{R@1k} \\
        \midrule
        \multirow{2}{*}{10K} & ANCE & 42.62 & 94.60 \\
         & + DAR & \textbf{46.31} & \textbf{94.81} \\
        \midrule
        \multirow{2}{*}{50K} & ANCE & 46.88 & \textbf{95.58} \\
         & + DAR & \textbf{48.20} & \textbf{95.58} \\
        \bottomrule
        \end{tabular}
        }
        \end{center}
        \vspace{-0.175in}
        \captionof{table}{\small Results on the MS MARCO subsets with ANCE as a denser retriever.}
        \label{tab:ance}
        \vspace{-0.2in}
    \end{minipage}
    \hfill
    \begin{minipage}{0.52\linewidth}
        \centering
        \begin{center}
        \resizebox{1.0\textwidth}{!}{
        \renewcommand{\arraystretch}{0.95}
        \begin{tabular}[h]{l|ccc}
        \toprule
        \multicolumn{1}{c|}{} &\textbf{Time (Min.)} &\textbf{Memory (MiB)} \\
        \midrule
        \multirow{1}{*}{DPR}
         & 19 & 22,071\\
        \multirow{1}{*}{DPR w/ QA}
         & 41 & 22,071\\
         \multirow{1}{*}{DPR w/ DA}
         & 38 & 22,071 \\
         \multirow{1}{*}{DPR w/ AR}
         & 29 & 38,986 \\
        \multirow{1}{*}{DAR (Ours)}
         & 21 & 22,071\\
        \bottomrule
        \end{tabular}
        }
        \end{center}
        \vspace{-0.175in}
        \captionof{table}{\small Wall-clock time and maximum memory usage for training a DPR model per epoch.}
        \label{tab:efficiency}
        \vspace{-0.2in}
    \end{minipage}
    \vspace{-0.05in}
\end{figure}


%% file: 0ablation_all.tex
\aboverulesep=0ex
\belowrulesep=0ex

\begin{figure*}[ht]
    \begin{minipage}{0.41\linewidth}
        \vspace{-0.18in}
        \centering
        \begin{center}
        \resizebox{1.0\linewidth}{!}{
        \renewcommand{\arraystretch}{1.05}
        \begin{tabular}[t]{l|cccccc}
        \toprule
        \multicolumn{1}{c|}{} &\textbf{MRR} &\textbf{MAP} &\textbf{T-20} &\textbf{T-5}\\
        \midrule
        \multirow{1}{*}{DAR (Ours)}
         & \textbf{42.92} & \textbf{27.12} & \textbf{75.04} & \textbf{57.62}\\
        \multirow{1}{*}{w/o Perturbation}
         & 41.26 & 26.19 & 73.68 & 55.37\\
        \multirow{1}{*}{w/o Interpolation}
         & 40.40 & 25.70 & 73.41 & 55.29\\
        \midrule
        \multirow{1}{*}{DPR}
         & 39.55 & 25.61 & 72.94 & 54.02\\
        \bottomrule
        \end{tabular}
        }
        \end{center}
        \vspace{-0.15in}
        \captionof{table}{\small Ablation studies of our DAR on the NQ dataset by removing interpolation or perturbation.}
        \label{tab:ablation}
    \end{minipage}
    \hfill
    \begin{minipage}{0.28\linewidth}
        \begin{center}
        \includegraphics[width=1.0\linewidth]{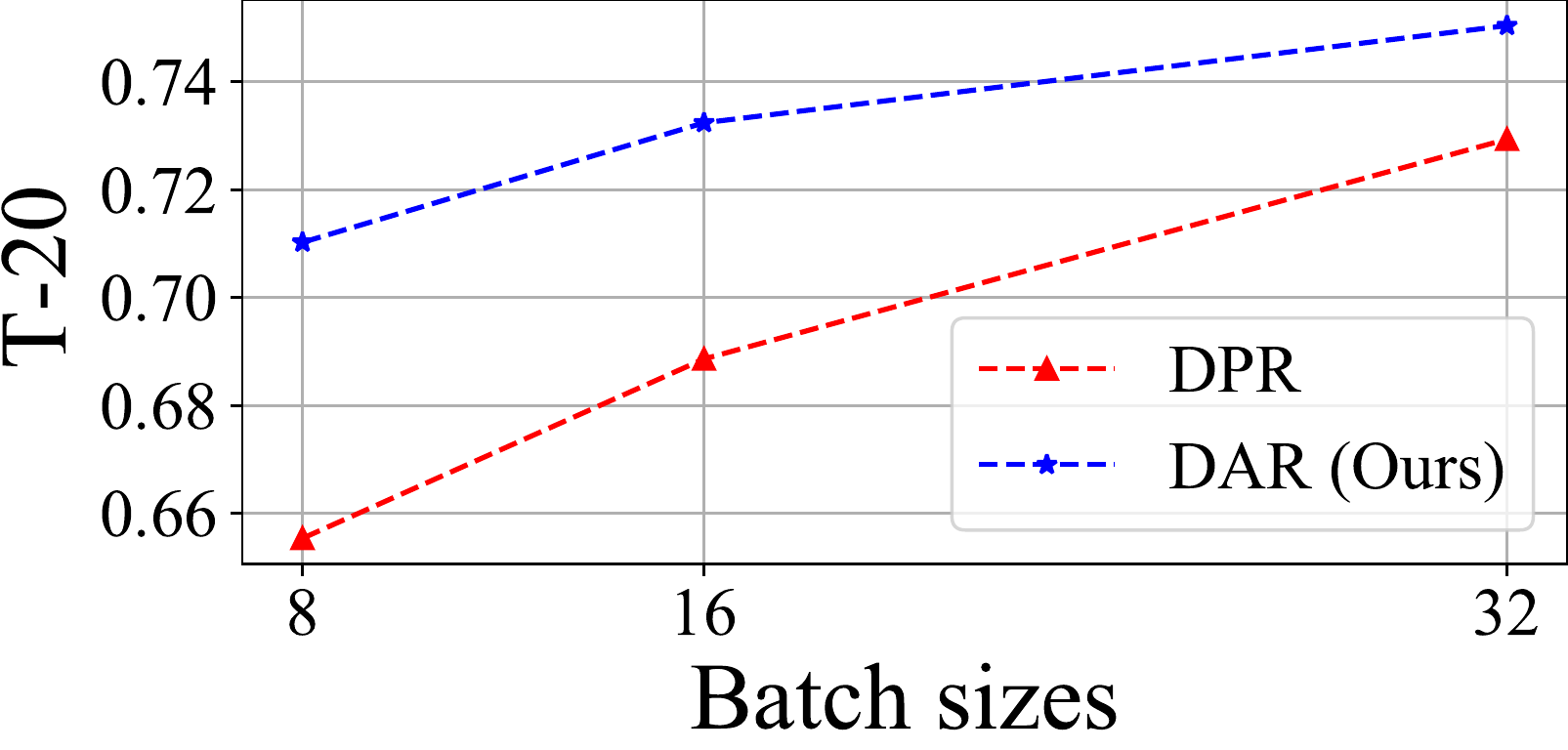}
        \end{center}
        \vspace{-0.2in}
        \caption{\small T-20 on the NQ dataset with varying batch sizes.
}
        \label{fig:batch_line}
    \end{minipage}
    \hfill
    \begin{minipage}{0.28\linewidth}
        \begin{center}
        \includegraphics[width=1.0\textwidth]{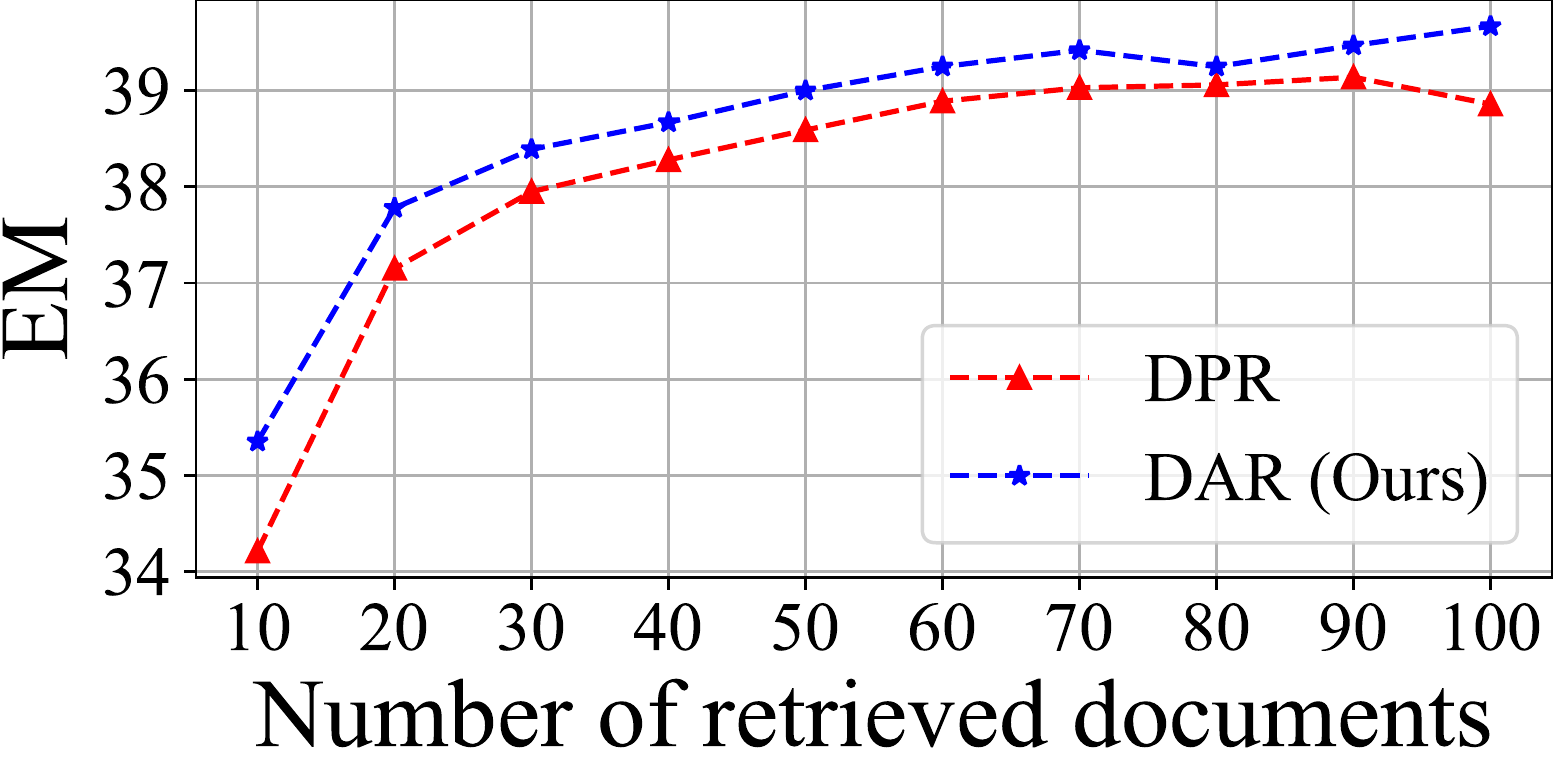}
        \end{center}
        \vspace{-0.2in}
        \caption{\small Exact Match (EM) scores for a reader on the NQ.
        }
        \label{fig:reader}
    \end{minipage}
    \vskip -0.175in
\end{figure*}

%% file: 4experiments.tex
\section{Experiments}

\subsection{Experimental Setups}
Here, we describe datasets, models, and implementation details for experiments. More experimental details are shown in \textbf{Appendix}~\ref{appendix:setup}. Our code is publicly available at \href{https://github.com/starsuzi/DAR}{github.com/starsuzi/DAR}.

\paragraph{Datasets}
For documents to retrieve, we use the Wikipedia, following~\citet{karpukhin2020dpr}, where the processed dataset contains 21,015,324 passages. To evaluate retrieval models, we use two open-domain QA datasets, following~\citet{karpukhin2020dpr}: 1) \textbf{Natural Questions (NQ)} is collected with Google search queries~\cite{Kwiatkowski2019NQ}; 2) \textbf{TriviaQA (TQA)} is a QA collection scraped from the Web~\cite{Joshi2017TriviaQAAL}.




\paragraph{Retrieval Models} \textbf{1) BM25} is a sparse term-based retrieval model based on TF-IDF~\cite{Robertson1994Okapi}. 
\textbf{2) Dense Passage Retriever (DPR)} is a dense retrieval model with a dual-encoder of query-document pairs~\cite{karpukhin2020dpr}. 
\textbf{3) DPR with Query Augmentation (DPR w/ QA)} augments pairs with query generation for the document, adopting~\cite{liang2020embedding, mao2021generation}.
\textbf{4) DPR with Document Augmentation (DPR w/ DA)} augments pairs by replacing words in the document~\cite{ma2019nlpaug}.
\textbf{5) DPR with Axiomatic Regularization (DPR w/ AR)} regularizes the retrieval model to satisfy certain axioms~\cite{axiomatic}.
\textbf{6) DAR} is ours with interpolation and perturbation of document representations.


\paragraph{Metrics}
\textbf{1) Top-K Accuracy (T-K)} computes whether a query's answer is included in Top-K retrieved documents. \textbf{2) Mean Reciprocal Rank (MRR)} and \textbf{3) Mean Average Precision (MAP)} measure the first rank and the average precision of query-relevant retrieved documents, respectively.


\paragraph{Implementation Details}
For the dense retrieval model based on the DPR framework, we refer to the publicly available code from DPR~\cite{karpukhin2020dpr}. We set the training epoch as 25 and batch size as 32 under academic budgets with a single GeForce RTX 3090 GPU having 24GB memory. We use in-batch negative sampling as our negative sampling strategy without hard negative samples. Also, we retrieve 100 passages per question. 

We use both interpolation and perturbation schemes for our augmentation methods. Specifically, for the interpolation method, we set $\lambda \in [0, 1]$ in equation~\ref{eq:interpolation} to be sampled from the uniform distribution. Also, for the perturbation method, we set the dropping rate as 0.1, and the number of dropout masks $n$ is selected in the range of 3 to 9.

\subsection{Results}
In this subsection, we show the overall performance of our DAR, and then give detailed analyses.

\vspace{-0.03in}
\paragraph{Overall Results}
As Table~\ref{tab:main} shows, DAR outperforms dense retrieval baselines on all datasets on the DPR framework. Note that DAR contributes to more accurate retrieval performance, since the smaller $K$ gives higher performance improvements. Furthermore, Figure~\ref{fig:seenunseen} shows that, with our method, the retrieval performance on unlabeled documents -- not seen during training -- together with the labeled ones is improved, where performance gains on unlabeled are remarkable. To see the robustness of DAR on other retrievers, we further evaluate our model on the recent ANCE framework (see \textbf{Appendix}~\ref{appendix:setup} for setups). As Table~\ref{tab:ance} shows, we observe that the performance improvement is more dominant on MRR when given 
a smaller number of training queries (low-resource settings), that DAR effectively augments document representations.

\vspace{-0.03in}
\paragraph{Results on Query Augmentation}
We focus on the problem of a notably small proportion of labeled documents in the training dataset, and propose to augment representations of unlabeled documents, which are not seen during training. However, it is also possible to augment representations of queries -- likely to be unseen at the test time -- by applying our interpolation and perturbation methods directly to queries. Note that we refer to our query augmentation method as Query Augmentation for dense Retrieval (QAR). As shown in Table~\ref{tab:main}, our proposed augmentation strategies also effectively improve the retrieval performance even when applied to queries. This result implies that our method is versatile, regardless of whether it is applied to documents or queries. 

\input{0negative_sampling}

\vspace{-0.03in}
\paragraph{Effectiveness of Interpolation \& Perturbation}
To understand how much our proposed interpolation and perturbation techniques contribute to the performance gain, we perform ablation studies. Table~\ref{tab:ablation} shows that each of the interpolation and stochastic perturbation positively contributes to the performance. In particular, when both of them are simultaneously applied, the performance is much improved, which demonstrates that these two techniques are in a complementary relationship.



\vspace{-0.03in}
\paragraph{Batch Size}
We test DAR with varying numbers of batch sizes. Figure~\ref{fig:batch_line} indicates that our DAR consistently improves the retrieval performance. Note that the smaller the batch size, the bigger the performance gap. Also, the batch size 16 of DAR outperforms the batch size 32 of the baseline, which highlights that DAR effectively augments document representations with a small batch.


\vspace{-0.03in}
\paragraph{Reader Performance}
To see whether accurately retrieved documents lead to better QA performance, we experiment with the same extractive reader from DPR without additional re-training. Figure~\ref{fig:reader} illustrates the effectiveness of our method on passage reading with varying numbers of retrieved documents. We observe that our retrieval result with small retrieved documents (i.e., $K=10$) significantly improves the performance of the reader. This implies that a more accurate retrieval on smaller $K$ in Table~\ref{tab:main} helps achieve the improved QA performance as~\citet{lee2021phrase} described. Furthermore, our reader performance may be further enhanced with advanced reading schemes~\cite{mao2021generation,qu2021rocketqa,ma2021reader}.


\vspace{-0.03in}
\paragraph{Negative Sampling Strategy}

To see the effectiveness of our DAR coupled with an advanced negative sampling scheme, we compare DAR against the baseline with the hard negative sampling strategy from BM25~\cite{karpukhin2020dpr}. Table~\ref{tab:negative_sampling} shows that DAR with hard negative sampling outperforms the baseline method. The results demonstrate that the performance of dense retrieval models could be further strengthened with a combination of our augmentation methods and advanced negative sampling techniques. Also, in all our experiments of the ANCE framework, we already use the strategy of negative sampling in~\citet{xiong2021approximate}, where we observe the clear performance improvement of our DAR on ANCE in Table~\ref{tab:ance}.

\input{0beir_table}

\vspace{-0.03in}
\paragraph{Results on Different Data Processing}

We additionally evaluate DAR on another NQ test dataset, following the processing procedure of~\citet{thakur2021beir}. For experiments, we reuse the same training checkpoint used in Table~\ref{tab:main}, as the training dataset is equal across the settings of~\citet{karpukhin2020dpr} and~\citet{thakur2021beir}. As Table~\ref{tab:beir} shows, our DAR also consistently outperforms all baselines when tested on the NQ test set from~\citet{thakur2021beir}. This confirms that our DAR robustly improves retrieval performances, regardless of the specific data processing strategies.


%% file: 0negative_sampling.tex
\aboverulesep=0ex
\belowrulesep=0ex

\begin{table}[t]
\centering
\begin{center}
\resizebox{0.48\textwidth}{!}{
\begin{tabular}[t]{l|cccccc}
\toprule
\multicolumn{1}{c|}{} &\textbf{MRR} &\textbf{MAP} &\textbf{T-100} &\textbf{T-1}\\
\midrule
\multirow{1}{*}{DPR+HN}
 & 53.40 & 33.38 & 84.82 & 43.21\\
\multirow{1}{*}{DAR+HN (Ours)}
 & \textbf{54.18} & \textbf{33.71} & \textbf{85.35} & \textbf{44.18}\\
\bottomrule
\end{tabular}
}
\end{center}
\vspace{-0.15in}
\caption{\small Retrieval results with hard negatives (HN) from BM25 on the NQ dataset for the DPR framework.}
\label{tab:negative_sampling}
\vspace{-0.2in}
\end{table}

%% file: 0beir_table.tex
\aboverulesep=0ex
\belowrulesep=0ex

\begin{table}[t!]
        \centering
        \begin{center}
        \resizebox{0.48\textwidth}{!}{
        \renewcommand{\arraystretch}{0.9}
        \begin{tabular}[t]{l|cccccc}
        \toprule
        \multicolumn{1}{c|}{} & \textbf{MRR} & \textbf{MAP} & \textbf{T-100} & \textbf{T-20} & \textbf{T-5} & \textbf{T-1} \\
        \midrule
        \multirow{1}{*}{BM25} & 29.60 & 28.05 & 77.87 & 61.30 & 42.27 & 18.86 \\
        \midrule
        \multirow{1}{*}{DPR} & 31.79 & 29.94 & 88.30 & 70.48 & 45.48 & 19.18 \\
        \multirow{1}{*}{DPR w/ QA} & 30.02 & 28.26 & 86.82 & 68.80 & 43.95 & 17.56 \\
        \multirow{1}{*}{DPR w/ DA} & 31.96 & 30.25 & 87.75 & 71.29 & 46.55 & 19.03 \\
        \multirow{1}{*}{DPR w/ AR} & 31.41 & 29.50 & 88.27 & 70.57 & 45.10 & 19.12 \\
        \multirow{1}{*}{DAR (Ours)} & \textbf{33.37} & \textbf{31.49} & \textbf{88.93} & \textbf{73.70} & \textbf{48.38} & \textbf{20.16}\\
        \bottomrule
        \end{tabular}
        }
        \end{center}
        \vspace{-0.15in}
        \caption{\small Retrieval results on the NQ dataset, following the processing procedure of~\citet{thakur2021beir}.}
        \label{tab:beir}
        \vspace{-0.175in}
\end{table}

%% file: 5conclusion.tex
\vspace{-0.05in}
\section{Conclusion}
\vspace{-0.1in}
We presented a novel method of augmenting document representations focusing on dense retrievers, which require an extensive amount of labeled query-document pairs for training. Specifically, we augment documents by interpolating and perturbing their embeddings with mixup and dropout masks. The experimental results and analyses on multiple benchmark datasets demonstrate that DAR greatly improves retrieval performances.

%% file: 6appendix.tex
\clearpage
\appendix
\section{Experimental Setups}
\label{appendix:setup}

\input{0qa_data_stat}

\paragraph{Datasets}
\label{appendix:setup:dataset}
To evaluate the performance of retrieval models, we need two types of datasets: 1) a set of documents to retrieve, and 2) pairs of a query and a relevant document, having an answer for the query. We first explain the datasets that we used for the DPR framework~\cite{karpukhin2020dpr}, and then describe the dataset for the ANCE framework~\cite{xiong2021approximate}.

For documents to retrieve, we use the Wikipedia snapshot from December 20, 2018, which contains 21,015,324 passages consisting of 100 tokens, following~\citet{karpukhin2020dpr} for the DPR framework. For open-domain QA datasets, we use Natural Question (NQ)~\cite{Kwiatkowski2019NQ} and Trivia QA (TQA)~\cite{Joshi2017TriviaQAAL}, following the dataset processing procedure of~\citet{karpukhin2020dpr}. We report the statistics of the training, validation, and test sets on NQ and TQA in Table~\ref{tab:qa_data_stat}.

To see the performance gain of our DAR on other dense retrieval models, we evaluate DAR on the ANCE framework~\cite{xiong2021approximate}, which is one of the recent dense retrieval models. ANCE is evaluated on the MS MARCO dataset, thus we use MS MARCO for training and testing our model. Note that training ANCE with the full MS MARCO dataset requires 225 GPU hours even after excluding the excessive BM25 pre-training and inference steps. Thus we randomly sample the MS MARCO dataset to train the model under academic budgets. Specifically, the subset of our MS MARCO passage dataset contains 500,000 passages. Also, we randomly divide the training queries into two subsets: one for 10,000 training queries and the other for 50,000 training queries. Then we align the sampled training queries to the query-document pairs in the MS MARCO dataset. On the other hand, we do not modify the validation set (dev set) of query-document pairs for testing. We summarize the statistics of the dataset in Table~\ref{tab:qa_data_stat}. Note that since the test set of MS MARCO is not publicly open, we evaluate the dense retrievers with the validation set, following~\citet{xiong2021approximate}.

\paragraph{Metrics}

\label{appendix:setup:metric}
Here, we explain the evaluation metrics for retrievers in detail. Specifically, given an input query, we measure the ranks of the correctly retrieved documents for the DPR framework with the following metrics:

\textbf{1) Top-K Accuracy (T-K):}
It measures whether an answer of the given query is included in the retrieved Top-K documents.

\textbf{2) Mean Reciprocal Rank (MRR):}
It computes the rank of the first correct document for the given query among the Top-100 retrieved documents, and then computes the average of the reciprocal ranks for all queries. 

\textbf{3) Mean Average Precision (MAP):}
It computes the mean of the average precision scores for all queries, where precision scores are calculated by the ranks of the correctly retrieved documents among Top-100 ranked documents. 

We use the following evaluation metric for the reader, which identifies the answer from retrieved documents.

\textbf{1) Exact Match (EM):}
It measures whether the reader exactly predicts one of the reference answers for each question.

Note that, for the ANCE framework, we follow the evaluation metrics, namely MRR@10 and Recall@1k, in the original paper~\cite{xiong2021approximate}.

\paragraph{Experimental Implementation Details}

For dense retrieval models based on the DPR framework, we follow the dual-encoder structure of query and document by using the publicly available code from DPR\footnote{https://github.com/facebookresearch/DPR}~\cite{karpukhin2020dpr}. For all experiments, we set the batch size as 32, and train models on a single GeForce RTX 3090 GPU having 24GB memory. Note that, in contrast to the best reported setting of DPR which requires industrial-level resources of 8 V100 GPUs (8 $\times$ 32GB = 256GB) for training with a batch size of 128, we use a batch size of 32 to train the model under academic budgets. We optimize the model parameters of all dense retrieval models with the Adam optimizer~\cite{KingmaB14adam} having a learning rate of 2e-05. We train the models for 25 epochs, following the analysis\footnote{See footnote 1.} that the training phases converge after 25 epochs.

For the retrievers based on the ANCE framework, we refer to the implementation from ANCE\footnote{https://github.com/microsoft/ANCE}~\cite{xiong2021approximate}. In order to directly measure the performance gain of the dense retrieval models based on ANCE from using our DAR, we use the pre-trained RoBERTa without warming up with the BM25 negatives. We train all the dense retrieval models for 50,000 steps with a single GeForce RTX 3090 GPU having 24GB memory, and simultaneously generate the ANN index with another GeForce RTX 3090 GPU, following~\citet{xiong2021approximate}. Following the standard implementation setting, we set the training batch size as 8, and optimize the model with the LAMB optimizer~\cite{You2020lamb} with a learning rate of 1e-6.

\paragraph{Architectural Implementation Details}
For our augmentation methods, we use both interpolation and perturbation schemes of document representations obtained from the document encoder $E_D$ in equation~\ref{eq:ir}. Specifically, given a positive query-document pair $(\boldsymbol{q}, \boldsymbol{d}^+)$, we first perturb the document representation $\boldsymbol{d}^+$ with dropout masks sampled from a Bernoulli distribution, which generates $n$ numbers of perturbed document representations $\left\{ \boldsymbol{d}^+_i \right\}_{i=1}^{i=n}$. Then, we augment them to generate $n$ numbers of positive query-document pairs $\left\{ (\boldsymbol{q}, \boldsymbol{d}^+_i) \right\}_{i=1}^{i=n}$, which we use in equation~\ref{eq:negative}. We search the number of perturbations $n$ in the range from 3 to 9, and set the probability of the Bernoulli distribution as 0.1.

Instead of only using positive or negative pairs, we further augment query-document pairs having intermediate similarities with mixup. Specifically, we interpolate representations between the perturbed-positive document $\boldsymbol{d}^+_i$ and the negative document $\boldsymbol{d}^-$ for the given query $\boldsymbol{q}$, with $\lambda \in [0, 1]$ in equation~\ref{eq:interpolation} sampled from a uniform distribution. Note that, given a positive pair of a query and a document, we consider the documents not identified as positive in the batch as negative documents. In other words, if we set the batch size as $32$, then we could generate $31$ interpolated document representations from $1$ positive pair and $31$ negative pairs. To jointly train the interpolation scheme with the original objective, we add the loss obtained from interpolation to the loss in equation~\ref{eq:negative}.

\section{Additional Experimental Results}
\subsection{Efficiency}
\label{appendix:results:efficiency}
As described in the Efficiency paragraph of Section~\ref{sec:method}, compared to the existing query augmentation methods~\cite{liang2020embedding,ma2021zero,qu2021rocketqa}, document augmentation method~\cite{ma2019nlpaug}, and word replacement method for regularization~\cite{axiomatic}, our method of augmenting document representations with interpolation and perturbation in a dense representation space is highly efficient. This is because, unlike the baselines above, we do not explicitly generate or replace a query or document text; but rather we only manipulate the representations of documents. This scheme greatly saves the time for training, since additional forwarding of the generated or replaced query-document pairs into the language model is not required for our data augmentation methods.

To empirically validate the efficiency of our methods against the baselines, we report the memory usage and time for training a retrieval model per epoch in Table~\ref{tab:efficiency}. As for memory efficiency, all the compared dense retrieval models using data augmentation methods, including ours, use the same amount of maximum GPU memory. This shows that the overhead of memory usage comes from operations in the large-size language model, such as BERT~\cite{devlin-etal-2019-bert}, not from manipulating the obtained document representations to augment the query-document pairs. Technically speaking, there are no additional parameters to augment document representations; thus our interpolation and perturbation methods do not increase the memory usage. On the other hand, DPR w/ AR excessively increases the memory usage, since it requires an extra forwarding process to the language model to represent the additional word-replaced sentences for regularization, instead of using the already obtained dense representations like ours.

We also report the training time for dense retrievers in Table~\ref{tab:efficiency}. Note that, for the explicit augmentation method based models, such as DPR w/ QA and DPR w/ DA, we exclude the extra time for training a generation model and generating a query or document for the given text. Also, we additionally generate the same number of query-document pairs in the training set, where the total amount of training data-points for DPR w/ QA and DPR w/ DA baselines are twice larger than the original dataset. Unlike these explicit query or document generation baselines, we perturb the document $n$ times, but also interpolate the representations of positive and negative documents. As shown in Table~\ref{tab:efficiency}, our DAR is about doubly more efficient than the explicit text augmentation methods, since DPR w/ QA and DPR w/ DA explicitly augment query-document pairs instead of using the obtained dense representations like ours. Also, our DAR takes a little more time to augment document representations than the base DPR model, while significantly improving retrieval performances as shown in Table~\ref{tab:main}. Even compared to the term replacement based regularization model (DPR w/ AR), our DAR shows noticeable efficiency, since an additional embedding process of the document after the word replacement on it requires another forwarding step besides the original forwarding step.

\input{0dpr_dev_table} 

\subsection{Reproduction of DPR}
We strictly set the batch size as 32 for training all the dense retrievers using the DPR framework; therefore the retrieval performances are different from the originally reported ones in~\citet{karpukhin2020dpr} that use a batch size of 128. However, while we use the available code from the DPR paper, one may wonder if our reproduction result is accurate. Therefore, since~\citet{karpukhin2020dpr} provided the retrieval performances of the DPR with different batch sizes (e.g., a batch size of 32), evaluated on the development (validation) set of the NQ dataset, we compare the Top-K accuracy between the reported scores and our reproduced scores. Table~\ref{tab:dpr_dev_table} shows that our reproduced Top-K accuracy scores with three different $K$s (e.g., Top-5, Top-20, and Top-100) are indeed similar to the reported ones, with ours even higher, thus showing that our reproductions are accurate.

\subsection{Experiment on WebQuestions}
One may have a concern that, as a sparse retrieval model -- BM25 -- outperforms all the other dense retrieval models on the TQA dataset in Table~\ref{tab:main}, TQA is not good enough to demonstrate the strength of our dense augmentation strategy. While we believe that sparse retrieval models are not our competitors as we aim to improve the dense retrieval models with data augmentation, in order to clear out such a concern, we additionally train and evaluate our DAR on the WebQuestions (WQ) dataset~\cite{webqa}, following the data processing procedure from~\cite{karpukhin2020dpr}. As Table~\ref{tab:webqa} shows, our DAR outperforms both dense and sparse retrieval models. Thus, the best scheme among sparse and dense retrievers still depends on the dataset, and combining sparse and dense models to complement each other will be a valuable research direction, which we leave as future work.
\input{0webqa_table}

%% file: 0qa_data_stat.tex
\aboverulesep=0ex
\belowrulesep=0ex

\begin{table}
\centering
\begin{center}
\resizebox{0.45\textwidth}{!}{
\begin{tabular}[h]{l|ccc}
\toprule
\multicolumn{1}{c|}{} &\textbf{Train} &\textbf{Val} &\textbf{Test} \\
\midrule
\multirow{1}{*}{Natural Question (NQ)}
 & 58,880 & 6,515 & 3,610 \\
\multirow{1}{*}{TriviaQA (TQA)}
 & 60,413 & 6,760 & 11,313 \\
\multirow{1}{*}{MS MARCO, \# Query: 10K}
 & 6,591 & 6,980 & - \\
 \multirow{1}{*}{MS MARCO, \# Query: 50K}
 & 32,927 & 6,980 & - \\
\bottomrule
\end{tabular}
}
\end{center}
\vspace{-0.15in}
\caption{\small Statistics for training, validation, and test sets on the NQ, TQA, and randomly sampled MS MARCO datasets. Note that, for MS MARCO, we only sample the number of training query-document pairs except for the validation set.}
\label{tab:qa_data_stat}
\vspace{-0.15in}
\end{table}

%% file: 0dpr_dev_table.tex
\aboverulesep=0ex
\belowrulesep=0ex

\begin{table}
    \centering
    \begin{center}
    \resizebox{0.46\textwidth}{!}{
    \renewcommand{\arraystretch}{0.95}
    \begin{tabular}[h]{l|ccc}
    \toprule
    \multicolumn{1}{l|}{} & \textbf{T-5} & \textbf{T-20} & \textbf{T-100} \\
    \midrule
     DPR \cite{karpukhin2020dpr}      & 52.1 & 70.8 & 82.1 \\
     DPR (Ours) & \textbf{53.2} & \textbf{71.6} & \textbf{82.7} \\
    \bottomrule
    \end{tabular}
    }
    \end{center}
    \vspace{-0.1in}
    \captionof{table}{\small Comparison of the DPR models' Top-K accuracy between the reported and reproduced scores. Best performance is highlighted in \textbf{bold}.}
    \label{tab:dpr_dev_table}
    \vspace{-0.15in}
\end{table}

%% file: 0webqa_table.tex
\aboverulesep=0ex
\belowrulesep=0ex

\begin{table}
\centering
\begin{center}
\resizebox{0.47\textwidth}{!}{
\renewcommand{\arraystretch}{0.9}
\begin{tabular}[ht]{l|cccccc}
\toprule
\multicolumn{1}{c|}{} &\textbf{MRR} &\textbf{MAP} &\textbf{T-100} &\textbf{T-20} &\textbf{T-5} &\textbf{T-1} \\
\midrule
\multirow{1}{*}{BM25}
 & 29.75 & 19.15 & 75.49 & 62.40 & 41.83 & 18.90 \\
\midrule
\multirow{1}{*}{DPR}
 & 33.34 & 21.76 & 78.64 & 65.75 & 45.87 & 22.00 \\
\multirow{1}{*}{DAR (Ours)}
 & \textbf{34.48} & \textbf{22.16} & \textbf{78.79} & \textbf{67.37} & \textbf{47.54} & \textbf{23.23} \\
\bottomrule
\end{tabular}
}
\end{center}
\vspace{-0.15in}
\caption{\small Retrieval results on the WQ dataset, in which the best performance is highlighted in \textbf{bold}.}
\label{tab:webqa}
\vspace{-0.2in}
\end{table}